\documentclass[onecolumn,preprintnumbers,amsmath,amssymb]{revtex4}
\usepackage{graphicx,subfigure,epsfig,epstopdf}
\usepackage{graphicx}% Include figure files
\usepackage{dcolumn}% Align table columns on decimal point
\usepackage{bm}% bold math
\usepackage{amsmath,amssymb}
\input{epsf}
\input{epsfx}
\newcommand{\unit}[1]{\hspace{-1.3pt} {#1}}

\begin{document}
\title{Terahertz Room-Temperature Photonic Crystal Nanocavity Laser}

\date{April 27, 2007}
\author{Dirk Englund}
\altaffiliation{Authors contributed equally.}
\affiliation{Ginzton Laboratory, Stanford University, Stanford CA 94305}
\author{Hatice Altug}
\altaffiliation{Authors contributed equally.}
\affiliation{Electrical and Computer Engineering Department, Boston University, Boston MA 02215}
\author{Ilya Fushman}
\affiliation{Ginzton Laboratory, Stanford University, Stanford CA 94305}
\author{Jelena Vu\v{c}kovi\'{c}}
\affiliation{Ginzton Laboratory, Stanford University, Stanford CA 94305}

%\section{Introduction}
\begin{abstract}
We describe an efficient surface-passivated photonic crystal nanocavity laser, demonstrating room-temperature operation with 3-ps total pulse duration (detector response limited) and low-temperature operation with an ultra-low threshold of 9\unit{$\mu$W}. 
\end{abstract}
%\pacs{42.50.Ct, 42.50.Dv, 42.70.Qs, 78.67.Hc}
%FIG1.pdf
\maketitle

%\section{Introduction}
Lasers in 2D photonic crystals (PC) hold great promise as single-mode light sources for low-power, high-speed applications in optical telecommunications, optical interconnects, and nano-scale sensing.  Their near-minimal mode volume allows for utilization of cavity quantum electrodynamic effects for improved threshold and speed \cite{Painter99science,Englund05PRL}. Furthermore, such enhancements can be achieved with even modest quality factor ($Q$) so that the cavity response time $\tau_p$ does not limit modulation rate.  We previously demonstrated a quantum well (QW) - driven coupled-cavity PC laser operating at cryogenic temperature and producing short pulses with rise time $\sim 1$\unit{ps}, fall-time $\sim 2$\unit{ps}, and FWHM$\sim 5$\unit{ps} \cite{Altug2006Nature}.  In this letter, we report on room-temperature operation with lasing response FWHM below $3$\unit{ps} and low-temperature continuous-wave (CW) operation with extremely low threshold power of 9\unit{$\mu$W}.  The operating temperature and threshold improvements are possible by increasing laser efficiency through surface passivation, while the speed-up results from faster carrier relaxation into the QW lasing level at room temperature. 

\begin{figure}%[htbp]
 \includegraphics[width=3in]{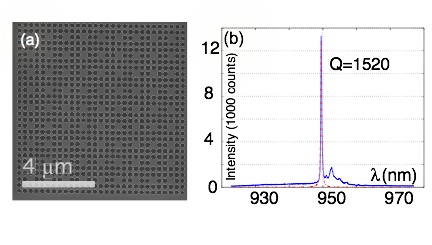} 
    \caption{Coupled cavity - photonic crystal laser. (a) Scanning electron micrograph of laser structure: 9x9 array of single-defect cavities.  Periodicity $a=315$\unit{nm}, hole radius $\sim 120$\unit{nm}, thickness 172 nm. (b) Lasing mode pumped at low power ($4 \mu$W, pulsed, pump laser diameter $\sim 5\mu$m). }
    \label{fig:Fig1}
\end{figure}

The lasers discussed here are similar to those described in \cite{Altug2006Nature}, consisting of 172 nm-thick GaAs slabs patterned with 9x9 arrays of coupled PC cavities in a square-lattice PC (Fig.\ref{fig:Fig1}). Four 8-nm In$_{0.2}$Ga$_{0.8}$As QWs separated by 8-nm GaAs barriers form the gain medium.  To reduce nonradiative (NR) surface recombination on the large QW area exposed through PC patterning, the sample was passivated with (NH$_4$)S, which resulted in a 3.7-fold reduction in the lasing threshold \cite{2007.APL.Englund}.  Measurements were obtained with a confocal microscope setup through a cryostat, as detailed in Ref.\cite{2007.OpEx.Englund}. 

The coupled cavity array laser is shown in Fig.\ref{fig:Fig1}(a). Only sets of small numbers of cavities in the array lase simultaneously as a result of fabriation imperfections. By spatially targeting the excitation laser at such sets, individual modes can be brought to lasing. Here we focus on single-mode lasing of a mode with $\lambda_{cav}=950$\unit{nm} at 10K (Fig.\ref{fig:Fig1}(b)).  The spontaneous emission (SE) rate enhancement in this resonant mode is estimated at $F_{cav}\approx 31$, following \cite{2007.APL.Englund}.  

\begin{figure}[htbp]
 \includegraphics[width=3in]{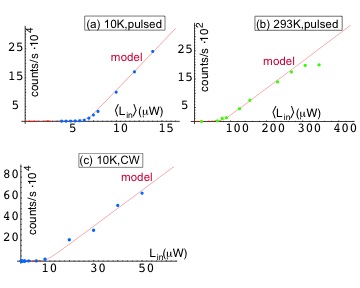}
    \caption{LL-curves.  (a) Low-temperature lasing (10K) with pulsed excitation (3.5\unit{ps}, 13\unit{ns}-rep.) shows an averaged threshold of 6.5$\mu$W, corresponding to $21$\unit{mW} peak power. (b) At room-temperature, lasing threshold increases to an averaged 68\unit{$\mu$W} (221\unit{mW} peak).  (c) Continous-wave lasing at low temperature shows a very low threshold of 9$\mu$W. Pump powers are measured before the objective lens.} % (d) For a different structure with $\sim 1$ cavities lasing and higher $Q\sim 3400$, the threshold only 2.2$\mu$W.  
    \label{fig:LLcurves}
\end{figure}

We first consider laser operation at low temperature (LT).  The following measurements were obtained at $10$K, though lasing operation does not change significantly up to $\sim 100$K. Fig.\ref{fig:LLcurves} (a) shows the lasing curve for pulsed excitation (3.5\unit{ps} at 13\unit{ns} repetition), with an averaged threshold of $6.5$\unit{$\mu$W} and corresponding peak power of $\sim 21$mW.  All powers reported are measured before the objective lens.  

\begin{figure}[htbp]
 \includegraphics[width=3in]{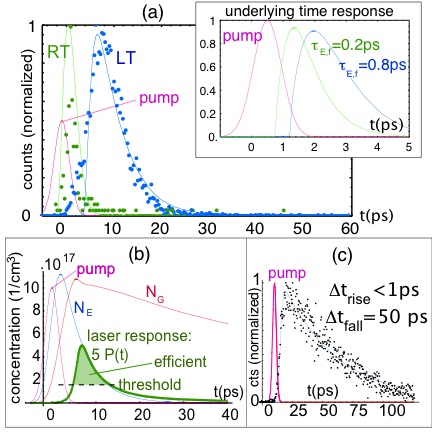}
    \caption{Photonic crystal laser time response.  (a) At low temperature (blue curve), FWHM=14.1\unit{ps} at $2 \times$ above lasing threshold.  At room-temperature (green), time response follows that of the pump laser with FWHM=$3.5$\unit{ps}.  \textit{Inset:} Calculated response to a shorter, 1-ps excitation pulse shows FWMH near 1 ps when pumped $2\times$ above threshold; response is faster if we assume faster carrier relaxation time $\tau_{E,f}$. (b) Illustration of pump inefficiency in pulsed operation.  Pump energy is efficiently channeled into the cavity mode only during lasing (shaded area under $P(t)$ curve, amplified here 5$\times$ for visibility); much of the remaining pump energy is wasted to SE and NR losses. (c)  Time-resolved PL from unpatterned QW at RT.  Faster response is possible through faster relaxation into the lasing level, estimated at $\tau_{E,f}<1$\unit{ps} from fitting to the rate model. }
    \label{fig:TR}
\end{figure}

CW lasing at LT shows a much lower threshold. Fig.\ref{fig:LLcurves}(c) displays the lasing curve, indicating onset of lasing at only $\sim 9$\unit{$\mu$W} CW pump power, considerably below other recently reported values for QW lasers\cite{2006.APL.Nomura,2006.APL.Shih}. Several factors contribute to the small threshold. One reason is that pulsed operation wastes pump energy when the laser mode periodically dips below threshold and carriers decay through inefficient SE or NR recombination, as illustrated in Fig.\ref{fig:TR}(b). Another reason is higher pump overlap with the active region as explained below.  

To quantify these contributions, we describe the laser action using the rate equations model given in Ref.\cite{2007.APL.Englund}.  The model considers carrier concentrations in the pump level ($N_E$) and lasing level ($N_G$), and the cavity photon density $P$ (number of photons in lasing mode divided by coupled cavity mode volume).  The pump level is excited with a laser at $780$\unit{nm}, above the GaAs bandgap. In Fig.\ref{fig:LLcurves}(a), we apply the model to the coupled nanocavity laser at LT in pulsed operation.  Here, the time-dependent photon density $P(t)$ is calculated following a $3.5$-ps Gaussian pump pulse and then averaged over the 13-ns repetition period to give the plotted output power. The parameters in the model are either directly measured or are standard values from literature\cite{endnote}; the only parameter we fit is the pump absorption efficiency $\eta=1.3\cdot 10^{-3}$ quantifying the fraction of the pump power that excites carriers in the pump level. The gain overlap appears to be lowered by carrier diffusion, which broadens the PL spot to $\sim 4\mu$m in diameter (as seen through the confocal microscope), considerably larger than the $\sim 1\mu$m pump laser spot size.  We note that although the model incorporates Auger recombination with coefficient $C_A=10^{-28}$cm$^{6}$/s \cite{1990HausserFuchs}, it is insignificant at present pump powers (it does become important at $\sim 10\times$ threshold power from an estimate of $N_G$). 
%
%XXXXXXX
%XXXX % .. we attribute the difference to the following two reasons: firstly, the PL is considerably smaller, probably because of less carrier diffusion in the material (that's b/c in CW, the average temperature is smaller, so there's less carrier diffusion, and explains why the spot size is smaller).  That give a factor 9 more carriers.  Secondly, directed carrier diffusion. In CW, we reach steady state, which allows more to be captured.  also say that th=thermal.  
In CW pumping, the PL spot size is considerably smaller, largely because a lower average thermal velocity $v_{th}$ of carriers leads to less diffusion.  From microscope imaging, we estimate it at $\sim 1.2\mu$m in diameter.  When this fact is taken into account in the above lasing model by $\eta\rightarrow 1.4\cdot 10^{-2}$ and otherwise identical parameters, we obtain a predicted threshold power of $L_{in,CW}=35\mu$W.  This value is still larger than the observed value of 9\unit{$\mu$W}.  We speculate that CW pumping is made more efficient by carrier drift into the lasing cavities: in the steady-state lasing regime, a carrier density gradient surrounding the coupled cavities channels carriers into them.  This effect exists in pulsed lasing as well, but is far less effective since the rate-limiting holes diffuse slowly compared to $\Delta \tau$, the fast lasing duration: $v_{th}\cdot \Delta \tau \sim 0.2\mu$m -- in other words, the cavities ceases to lase before a significant fraction of carriers could have diffused into them.  We estimate that it is reasonable that this carrier drift in the CW regime improves the pump absorption efficiency to $\eta \rightarrow 0.055$ as the cavities effectively capture $\sim 4\times$ more carriers.  The model then describes the CW lasing curves well, as shown in Fig.\ref{fig:LLcurves}(c).  Thus, the remarkably low CW threshold appears possible through three primary contributions: efficient conversion of carriers into lasing mode photons in steady-state lasing, a smaller PL spot size, and carrier drift into the lasing mode. %The low CW threshold observed in this structure was not uncommon; a neighboring structure on the same chip showed an even lower threshold of only $2.2\mu$W.  Adjusting for that mode's smaller size, about one coupled cavities judging from the microscope image, and higher $Q\approx 3400$, the model again fits the data well (Fig.\ref{fig:LLcurves},d).

The measurements described so far were obtained at low temperature. Room-temperature (RT) operation is more challenging because of heating problems associated with higher threshold, and was previously not possible with our structures. We achieved RT lasing after suppressing NR surface recombination using a surface passivation technique\cite{2007.APL.Englund}. We first consider the easier case of RT lasing in the pulsed mode.  The lasing curve in Fig.\ref{fig:LLcurves}(b) indicates a threshold of 68\unit{$\mu$W} averaged power. The larger threshold results in part from higher transparency concentration and smaller optical gain \cite{1995Coldren}.  Threshold is also roughly proportional to the NR surface recombination rate, which is as fast or faster than the SE rate in this type of PC structure\cite{2007.APL.Englund}.  From separate lifetime measurements on bulk and patterned QW regions, we estimate the NR recombination lifetime to drop from $188$\unit{ps} at LT to $50$\unit{ps} at RT. In addition to pulsed operation, we also achieved quasi-CW operation at RT.  This required a chopper wheel that provided 1\unit{ms}-long pulses at a 17 Hz repetition rate. However, operation was too transient to measure a LL curve reliably.  

RT operation allows remarkably fast full-signal laser modulation rates.  In Fig.\ref{fig:TR}(a), we present streak camera measurements of the lasing response to 3.4-ps-long pump pulses (13\unit{ns} repetition) at LT and RT.  Both measurements were obtained with pump powers roughly $2\times$ above threshold, corresponding to averaged pump powers of 13\unit{$\mu$W} and 136$\mu W$ at LT and RT, respectively.  We measured significantly faster lasing response at RT, with the lasing pulses roughly following the pump duration.  This speed-up is due to faster phonon-mediated carrier relaxation at RT, as indicated in the PL response from the unpatterned QW at RT, shown in Fig.\ref{fig:TR}(c):  the rise-time $\tau_{E,f}$ is streak-camera limited to less than 1\unit{ps}, significantly shorter than the LT rise-time of $\sim 6$\unit{ps}.  This behavior is captured well by the three-level rate equations model whose calculated response is convolved with a filter that takes into account the 3.2-ps response time (FWHM) of the streak camera \cite{2007.APL.Englund}.  Based on our model, lasing response should approach FWHM=$1.2$\unit{ps} at 2$\times$ threshold pump power when pumped with shorter $1$-ps laser pulses, implying modulation rates in the THz regime.  This is shown in Fig.\ref{fig:TR}(a, inset), where the lasing response is modeled for two values of the carrier relaxation time into the lasing level,  $\tau_{E,f}=0.8$ps and $\tau_{E,f}=0.2$ps.  The delay can be decreased with increasing pump power, but is ultimately limited by the carrier relaxation time $\tau_{E,f}$.

%To achieve more stable operation, we surround the PC structure in low-index material for increased thermal dissipation.  The GaAs membrane was surrounded by an 800-\unit{nm}-layer of oxidized Al$_0.9$Ga$_0.1$As on the bottom or with a 500-\unit{nm}-thick Poly(methyl methacrylate) (PMMA) layer on top, or combinations of both. The PMMA layer served the additional purpose of capping the surface to prevent renewed surface passivation.  In all combinations, thermal stability was increased dramatically, allowing $\sim 20\times$ higher pumping power without damage to the membrane. We presently we find that $Q$ values are reduced significantly to below $300$, so stable CW operation at RT has not yet been achieved.  We believe that this problem arises from addressable fabrication problems; FDTD simulations show that the capping layers should not reduce $Q$ below $Q<4000$. 
In conclusion, we have described the laser dynamics in a surface-passivated GaAs PC coupled nanocavity laser employing an InGaAs quantum well gain medium.  At low temperature, we observe remarkably low CW threshold pump powers near 9\unit{$\mu$W}.  At room temperature, increased surface recombination, higher transparency concentration, and lower QW gain increase the threshold and lead to heating problems.  Thus we only observe transient CW operation.  To address this problem, we are currently investigating sandwiching the PC membrane between PMMA and/or oxidized Al$_{0.9}$Ga$_{0.1}$As, which improves heat exchange by $\sim 20\times$ while permanently capping the structure to prevent re-oxidation. On the other hand, pulsed operation at room-temperature is very stable. The increased relaxation rate into the lasing level at room temperature enables very fast modulation rates; we observe laser pulses with FWHM near the 3.4-ps pump-pulse duration, with rise-and fall times $\sim$ 1 ps.  Our three-level laser model agrees well with experimental observations and indicates the PC laser has nearly cavity-lifetime-limited response time, putting our structures in the THz modulation rate regime.
% for the cavity-coupled array, and threshold powers as low as $2.2\mu$W for single defect cavities

This work was supported by the MARCO IFC Center, NSF Grants ECS-0424080 and ECS-0421483, the MURI Center  (ARO/DTO Program No.DAAD19-03-1-0199), as well as NDSEG Fellowships (D.E. \& I.F.).  
\newpage
\clearpage

%\bibliographystyle{unsrt}
%\bibliography{../../../latex/references}

%(b) At high pump power (40mW, pulsed, laser focus $\sim 30\mu$m), all lasing modes are visible. (

\end{document}